\documentclass[12pt]{article}
\usepackage{amsfonts}
\usepackage{amssymb}

\def\hybrid{\topmargin -20pt    \oddsidemargin 0pt
        \headheight 0pt \headsep 0pt
       \textwidth 6.35in       
        \textheight 9.25in       

       \marginparwidth .875in
        \parskip 5pt plus 1pt   \jot = 1.5ex}

\hybrid

\catcode`\@=11

\def\marginnote#1{}
\newcount\hour
\newcount\minute
\newtoks\amorpm
\hour=\time\divide\hour by60
\minute=\time{\multiply\hour by60 \global\advance\minute by-\hour}
\edef\standardtime{{\ifnum\hour<12 \global\amorpm={am}%
        \else\global\amorpm={pm}\advance\hour by-12 \fi
        \ifnum\hour=0 \hour=12 \fi
        \number\hour:\ifnum\minute<10 0\fi\number\minute\the\amorpm}}
\edef\militarytime{\number\hour:\ifnum\minute<10 0\fi\number\minute}

\def\draftlabel#1{{\@bsphack\if@filesw {\let\thepage\relax
   \xdef\@gtempa{\write\@auxout{\string
      \newlabel{#1}{{\@currentlabel}{\thepage}}}}}\@gtempa
   \if@nobreak \ifvmode\nobreak\fi\fi\fi\@esphack}
        \gdef\@eqnlabel{#1}}
\def\@eqnlabel{}
\def\@vacuum{}
\def\draftmarginnote#1{\marginpar{\raggedright\scriptsize\tt#1}}

\def\draft{\oddsidemargin -.5truein
        \def\@oddfoot{\sl preliminary draft \hfil
        \rm\thepage\hfil\sl\today\quad\militarytime}
        \let\@evenfoot\@oddfoot \overfullrule 3pt
        \let\label=\draftlabel
        \let\marginnote=\draftmarginnote
   \def\@eqnnum{(\theequation)\rlap{\kern\marginparsep\tt\@eqnlabel}%
\global\let\@eqnlabel\@vacuum}  }

\def\preprint{\twocolumn\sloppy\flushbottom\parindent 2em
        \leftmargini 2em\leftmarginv .5em\leftmarginvi .5em
        \oddsidemargin -.5in    \evensidemargin -.5in
        \columnsep .4in \footheight 0pt
        \textwidth 10.in        \topmargin  -.4in
        \headheight 12pt \topskip .4in
        \textheight 6.9in \footskip 0pt
        \def\@oddhead{\thepage\hfil\addtocounter{page}{1}\thepage}
        \let\@evenhead\@oddhead \def\@oddfoot{} \def\@evenfoot{} }

\def\numberbysection{\@addtoreset{equation}{section}
        \def\theequation{\thesection.\arabic{equation}}}

\def\underline#1{\relax\ifmmode\@@underline#1\else
        $\@@underline{\hbox{#1}}$\relax\fi}

\def\titlepage{\@restonecolfalse\if@twocolumn\@restonecoltrue\onecolumn
     \else \newpage \fi \thispagestyle{empty}\c@page\z@
        \def\thefootnote{\fnsymbol{footnote}} }

\def\endtitlepage{\if@restonecol\twocolumn \else \newpage \fi
        \def\thefootnote{\arabic{footnote}}
        \setcounter{footnote}{0}}  

\catcode`@=12
\relax

\def\figcap{\section*{Figure Captions\markboth
        {FIGURECAPTIONS}{FIGURECAPTIONS}}\list
        {Figure \arabic{enumi}:\hfill}{\settowidth\labelwidth{Figure
999:}
        \leftmargin\labelwidth
        \advance\leftmargin\labelsep\usecounter{enumi}}}
 \relax
\def\tablecap{\section*{Table Captions\markboth
        {TABLECAPTIONS}{TABLECAPTIONS}}\list
        {Table \arabic{enumi}:\hfill}{\settowidth\labelwidth{Table
999:}
        \leftmargin\labelwidth
        \advance\leftmargin\labelsep\usecounter{enumi}}}
 \relax
\def\reflist{\section*{References\markboth
        {REFLIST}{REFLIST}}\list
        {[\arabic{enumi}]\hfill}{\settowidth\labelwidth{[999]}
        \leftmargin\labelwidth
        \advance\leftmargin\labelsep\usecounter{enumi}}}
 \relax
\makeatletter
\newcounter{pubctr}
\def\publist{\@ifnextchar[{\@publist}{\@@publist}}
\def\@publist[#1]{\list
        {[\arabic{pubctr}]\hfill}{\settowidth\labelwidth{[999]}
        \leftmargin\labelwidth
        \advance\leftmargin\labelsep
        \@nmbrlisttrue\def\@listctr{pubctr}
        \setcounter{pubctr}{#1}\addtocounter{pubctr}{-1}}}
\def\@@publist{\list
        {[\arabic{pubctr}]\hfill}{\settowidth\labelwidth{[999]}
        \leftmargin\labelwidth
        \advance\leftmargin\labelsep
        \@nmbrlisttrue\def\@listctr{pubctr}}}
 \relax
\makeatother
\newskip\humongous \humongous=0pt plus 1000pt minus 1000pt

\newif\ifdtup

\relax

\def\be{\begin{equation}}
\def\ee{\end{equation}}
\def\ba{\begin{eqnarray}}
\def\ea{\end{eqnarray}}

\def\del{\partial}

\def\r{\rho}
\def\a{\alpha}

\def\b{\beta}

\def\g{\gamma}
\def\G{\Gamma}
\def\d{\delta}
\def\D{\Delta}
\def\e{\epsilon}

\def\m{\mu}
\def\n{\nu}
\def\om{\omega}
\def\Om{\Omega}
\def\l{\lambda}
\def\L{\Lambda}
\def\s{\sigma}
\def\S{\Sigma}

\def\cN{{\cal N}}

\def\cL{{\cal L}}

\def\bs{\bigskip}
\def\no{\noindent}

\def\qq{\qquad}

\def\IR{\relax{\rm I\kern-.18em R}}

\def \ha {{1\over 2}}

\def \ov {\over}

\def\const{{\rm const.}}

\def\IR{\relax{\rm I\kern-.18em R}}
\def\inv{^{\raise.15ex\hbox{${\scriptscriptstyle -}$}\kern-.05em 1}}

\def\cL{{\cal L}}

\def\tr{{\rm Tr}}

\def\hi{{\hat i}}
\def\hj{{\hat j}}
\def\hk{{\hat k}}

\begin{document}


\renewcommand{\theequation}{\thesection.\arabic{equation}}

\newcommand{\beq}{\begin{equation}}
\newcommand{\eeq}[1]{\label{#1}\end{equation}}
\newcommand{\ber}{\begin{eqnarray}}
\newcommand{\eer}[1]{\label{#1}\end{eqnarray}}
\newcommand{\eqn}[1]{(\ref{#1})}
\begin{titlepage}
\begin{center}

\hfill NEIP-01-012\\
\vskip -.05 cm
\hfill hep--th/0112117\\
\vskip -.05 cm
\hfill November 2001

\vskip .5in
\baselineskip 17.5pt
{\large \bf Integrable reductions of $Spin(7)$ and $G_2$ invariant self-dual
Yang--Mills equations and gravity}

\vskip 0.4in

{\bf Konstadinos Sfetsos}
\vskip 0.07in
{
Institut de Physique, Universit\'e de Neuch\^atel\\
Breguet 1, CH-2000 Neuch\^atel, Switzerland\\
\vskip 0.1in
and\\
\vskip 0.1in
Department of Engineering Sciences,
University of Patras\\
26110 Patras, Greece\\
\footnotesize{\tt sfetsos@mail.cern.ch}}\\

\end{center}

\vskip .3in

\centerline{\bf Abstract}
\baselineskip 17.5pt
\noindent
There is remarkable relation between self-dual Yang--Mills
and self-dual Einstein gravity in four Euclidean dimensions.
Motivated by this we investigate
the $Spin(7)$ and $G_2$ invariant self-dual Yang--Mills equations
in eight and seven Euclidean dimensions and search for their possible
analogs in gravitational theories. The reduction of the self-dual Yang--Mills
equations to one dimension results into systems of first order
differential equations.
In particular, the $Spin(7)$-invariant case gives rise to a
7-dimensional system which is completely integrable.
The different solutions are classified
in terms of algebraic curves and are characterized by the genus of
the associated Riemann surfaces. Remarkably, this system arises also
in the construction of solutions in gauged supergravities that have an
interpretation as continuous distributions of branes in string and M-theory.
For the $G_2$ invariant case we perform two distinct reductions, both
giving rise to 6-dimensional systems.
The first reduction, which is a
complex generalization of the 3-dimensional Euler spinning top system,
preserves an $SU(2)\times SU(2)\times Z_2$ symmetry and is fully integrable in
the particular case where an extra $U(1)$ symmetry exists.
The second reduction we employ, generalizes the Halphen system
familiar from the dynamics of monopoles.
Finally, we analyze massive generalizations and present solitonic solutions
interpolating between different degenerate vacua.

\vskip .4in
\noindent

\end{titlepage}
\vfill
\eject

\def\baselinestretch{1.2}
\baselineskip 17.5pt
\noindent


\section{Introduction}
\setcounter{equation}{0}

In an important paper almost two decades ago, the self-duality equations
for the gauge field strength of Yang--Mills (YM) theories
were examined in a general $D$-dimensional
Euclidean space \cite{Corrigan}. These authors
classified all possible cases for $D=5,6,7,8$ in terms of the
maximal subgroups of the rotation group $SO(D)$ that leave invariant
a 4-index totally antisymmetric tensor.
The eight-dimensional case with invariance subgroup the $Spin(7)$ of $SO(8)$
is quite distinct
in the sense that it generalizes more closely the four-dimensional
self-duality. The work of \cite{Corrigan}
was based on group theory, but
nevertheless, the analogy with four-dimensional self-dual YM equations
was pushed further when the eight-dimensional
analogue of the four-dimensional instanton solution was found \cite{Fubini}.
This solution can also be embedded in heterotic string theory
\cite{HarveyStro}.
The seven-dimensional case with invariance subgroup the $G_2$ of $SO(7)$
is also quite interesting and can be discussed in parallel with the
eight-dimensional $Spin(7)$ case \cite{Corrigan}. Also in this case
a seven-dimensional instanton solution was found, as well as its
embedding in heterotic string theory \cite{GunaNi} (see also
\cite{ivanova}).

In this paper we reexamine in detail the self-dual YM equations in eight and
seven dimensions with the invariance groups $Spin(7)$ and $G_2$, respectively,
that we mentioned. In particular, we consider
the systems arising from reducing them to one dimension,
when all fields depend only on one space variable.
Our motivation to perform this kind of reduction stems from the fact
that, in a similar setting in four Euclidean dimensions,
there is a remarkable relation between self-dual YM and self-dual gravity.
We recall that the four-dimensional self-dual YM equations
\be
F_{\m\n}=\ha \e_{\m\n\r\s}F_{\r\s}\ ,
\ee
with the gauge choice $A_4=0$ and in the ansatz that the remaining fields
$A_i$, $i=1,2,3$ depend only on the variable $x_4\equiv \tau$, reduce to
the Nahm system
\be
{dA_i\ov d\tau}=\ha \e_{ijk} [A_j,A_k]\ ,
\label{nahmm}
\ee
which appeared in the theory of static non-abelian monopoles \cite{Namo}.
Choosing $SU(2)$ as the gauge group one can parametrize the gauge fields as
$A_j=-i\om_j\s_j/2$ (no sum over $j$),
where the $\s_i$'s are Pauli matrices and
derive the system of equations (see, for instance, \cite{Chakravarty})
\be
{d\om_1\ov d\tau}=\om_2 \om_3\ ,\quad ({\rm and\ cyclic\ perms.}) \ .
\label{eell}
\ee
This is the well known Lagrange system that also coincides with the Euclidean
continuation of the
three-dimensional Euler top equations
describing the free motion of a rigid body
with one point fixed.
There is a inequivalent parametrization
of the gauge field that results instead of \eqn{eell} to \cite{Chakravarty,Takhtajan}
\be
{d\om_1\ov d\tau}=-\om_2 \om_3 + \om_1(\om_2+\om_3)\ ,\quad
({\rm and\ cyclic\ perms.}) \ ,
\label{eell1}
\ee
which is known as the Halphen system.
A well known remarkable fact is that, both of the above systems
also arise in self-dual
four-dimensional Einstein gravity with metrics having an $SO(3)$ isometry.
As famous examples we mention, the Egutchi--Hanson metric \cite{EgH},
described by the
system \eqn{eell} and the Taub--NUT and Atiyah--Hitchin metrics \cite{athi}
described by \eqn{eell1} \cite{Gipo}.

Recently, there is quite a bit of interest on special holonomy
$Spin(7)$ and $G_2$, metrics
of eight- and seven-dimensional Einstein gravity due to their relevance
not only in mathematics, but also in physics (see, for instance,
\cite{Acharya}-\cite{AtWi}).
In practice, the computation reduces mathematically to solving a system
of first order ordinary differential equations of the type in \eqn{eell} and
\eqn{eell1}, but typically much more complicated since in general they
involve six or seven unknown functions.
Although there are a few scattered important solutions
\cite{Bryant}-\cite{Gukov}
no systematic
study of these systems exists in the literature and finding new solutions
proves a very difficult task.
We believe that making connection with the eight- or
seven-dimensional self-dual YM equations
is a promising avenue for such a systematic approach towards
their integrability. In fact, there is at least one case where this
correspondence in precise and complete.
We will see that
for the $Spin(7)$ case the resulting 7-dimensional system coincides with one
that arose before in the construction of domain wall solutions
in gravity coupled to scalars theories \cite{bs1,bs2}.
These theories are sectors of gauged supergravities in four, five
and seven dimensions and the domain wall solutions, when viewed from a
string or an M-theory point of view, represent the gravitational field of
a large number of continuously distributed $p$-branes.
As such this system
turns out to be completely integrable with the help of algebraic
curves and the various solutions are characterized by the genus of
associated Riemann surfaces.

The rest of the paper is organized as follows: In section 2 we consider
the $Spin(7)$ invariant eight-dimensional self-dual YM equations dimensionally
reduced to one space dimension.
We present the general solution in terms of an
auxiliary function and reduce the problem
mathematically to the study of a non-linear differential
equation this function obeys.
In section 3 we
give a number of elementary, albeit non-trivial, examples. In section 4
we associate the differential equation that we mentioned above with algebraic
curves and the corresponding Riemann surfaces which then are used
in order to classify all inequivalent solutions. We also present our
main results in tables and make the precise connection with
solutions of gauged supergravity in various dimensions and their lift to
M- and string theory.
In section 5 we consider a massive extension
of the $Spin(7)$ self-dual invariant eight-dimensional self-dual YM equations.
We study the vacuum structure of this theory and show that there are
solitonic solutions interpolating between degenerate isolated vacua.
In the simplest case we obtain the usual kink solution
in a theory of one scalar self-interacting with a ``mexican
hat'' potential, but we also exhibit other examples.
In section 6 we consider
the $G_2$ invariant seven-dimensional self-dual YM equations again
reduced to one dimension.
We perform two distinct reductions and obtain six-dimensional systems
of differential
equations that are complex generalizations of the Lagrange and Halphen systems
\eqn{eell} and \eqn{eell1} above. We provide constants on motion and in one
particular case the full solution. We also present a massive generalization
based on the analogy we develop with weak $G_2$ holonomy metrics.
We end the paper with a few concluding remarks and some feature directions of
this work in section 7. We have also written an appendix with some useful
properties of the octonionic structure constants and related tensors.

\section{8D self-dual YM with $Spin(7)$ invariance}
\setcounter{equation}{0}

Consider the eight-dimensional self-duality equations \cite{Corrigan}
\be
F_{\a\b}=\l \Psi_{\a\b\g\d} F_{\g\d}\ .
\label{sellf}
\ee
where $\a=1,2,\dots, 8$ and the gauge field strength is
$F_{\a\b}=\del_a A_{\b}-\del_{\b} A_{\a} - [A_{\a},A_{\b}]$.
The totally antisymmetric 4-index tensor $\Psi_{\a\b\g\d}$ is invariant
under the $Spin(7)$ subgroup of the rotational group $SO(8)$.
Its components are constructed in terms of the structure constants of the
octonionic algebra. Some useful properties of these tensors
are collected in the appendix. Note also that, solutions of the
self-duality equations \eqn{sellf} automatically provide solutions
to the
equations of motion for the gauge fields, since the latter
are reduced to the Bianchi identity
due to the antisymmetry of the tensor $\Psi_{\a\\b\g\d}$.
As already noted in \cite{Corrigan}, consistency
of \eqn{sellf} requires either one of the four values
\be
\l={\e\ov 2}\ ,\qq \l=-{\e\ov 6}\ ,\qq \e=\pm 1\ .
\label{lllco}
\ee
We pick a particular
direction, say the eighth, and split the index $\a=(a,8)$, where
the $a=1,2,\dots , 7$. We break the 28 independent conditions in
\eqn{sellf} into
\be
F_{a8}=\l \psi_{abc} F_{bc}\ ,
\label{sell1}
\ee
representing $7$ conditions and
\be
F_{ab}=2 \l \psi_{abc} F_{c8} + \l \psi_{abcd} F_{cd}\ ,
\label{sell2}
\ee
representing the remaining $21$ conditions.
One can show that the 21 conditions in \eqn{sell2} imply the 7 conditions in
\eqn{sell1} if the parameter $\l$ takes either one of the four values in
\eqn{lllco}. However, the 7 conditions in \eqn{sell1} imply the 21 conditions
in \eqn{sell2} only for the value $\l=\e/2$.

In the rest of the paper we
restrict to the self-dual case with $\l=\ha$ (the anti-self-dual
case with $\l=-\ha$ is recovered trivially) which means to the ${\bf 21}$ of
$Spin(7)$. Then, solving
the system of 7 equations in \eqn{sell1}
we automatically provide a solution to \eqn{sellf} (for $\l=\ha$).
We next
make the gauge choice $A_8=0$ and we look for solutions that depend only
on the eighth coordinate $x^8\equiv \tau$. Then \eqn{sell1} (for $\l=\ha$)
becomes
\be
{d A_a\ov d\tau}= \ha \psi_{abc} [A_b,A_c]\ ,\qq a,b,c=1,2,\dots , 7\ .
\label{om122}
\ee
which is the seven-dimensional generalization of the Nahm system \eqn{nahmm}.
Writing $A_a=\psi_a \om_a$ (no sum over $a$),
where $\psi_i$ is the adjoint-like
representation, defined in \eqn{addj}, we obtain the system of 7 coupled
non-linear equations
\be
{d\om_a\ov d\tau}= \ha \psi^2_{abc} \om_b \om_c\ ,\qq a,b,c=1,2,\dots ,7\ ,
\label{om12}
\ee
where we have used the properties \eqn{addj1} and \eqn{addj2}.
This is the generalization of the three-dimensional Lagrange system
\eqn{eell} to seven dimensions and it was first obtained in
\cite{Floratos,FaiUeno}.\footnote{It is interesting that
solutions of \eqn{om12} can be used to construct solutions of the
self-dual membrane embedded in 8-dimensions \cite{Floratos}.}
These first order equations imply the second order ones
\be
{d^2 \om_a\ov d\tau^2}=
\ha \psi^2_{abc} \psi^2_{bde} \om_c \om_d \om_e\ .
\ee
The system \eqn{om12} represents a flow in a seven-dimensional manifold
spanned by the $\om_a$'s. Moreover, it is a gradient flow in
the sense that there exists a prepotential $W$
such that the first order system is obtained as
\be
{d\om_a\ov d\tau}= g_{ab} {\del W\ov \del\om_b}\ ,\qq a,b=1,2,\dots ,7\ ,
\label{grom}
\ee
for some metric $g_{ab}$ in the space of the $\om_a$'s.
The corresponding second order equations can be derived from the Lagrangian
\be
\cL = -\ha g\inv_{ab}\dot\om_a\dot \om_b - V\ ,
\label{laab2}
\ee
with the potential given in terms of the prepotential $W$ as
\be
V=\ha g_{ab} {\del W\ov \del\om_a}{\del W\ov \del\om_b}\ .
\label{grv}
\ee
Every solution to the first order system solves the second order
Lagrange equations for \eqn{laab2}. However, the reverse is of course
not true. Namely, not every solution to the second order
Lagrange equations satisfies the first order system \eqn{grom}.
In general, a prepotential has a number of critical points for the $\om_a$'s
which are
found by solving the algebraic system of equations ${\del W/\del \om_a}=0$.
For a positive definite metric $g_{ab}$ every critical point of the
prepotential corresponds to a minimum $V=0$ for the potential. Other extrema
the potential itself might have, necessarily correspond to its maxima.
In our case the prepotential and the metric are given by
\be
W =  {1 \ov 6} \psi_{abc}^2 \om_a \om_b \om_c \ ,\qq g_{ab} = \d_{ab} \ .
\label{hh2}
\ee

\subsection{Solving the first order equations}

In order to solve \eqn{om12} it is convenient to make the standard
choice for the set of structure constants $\psi_{abc}$ given by \eqn{g2str1}.
Let's next define a new set of variables as
\be
\Om_a=M_{ab} \om_b\ ,
\label{ttrr}
\ee
where the matrix $M$ and its inverse are given by
\be
M=\pmatrix{0& 1 & 1 & 1 & 0 & 0 & 1\cr
1& 0& 1& 1& 0& 1& 0\cr
0 & 0 & 0& 1 & 1 & 1 & 1\cr
1 & 1 & 0 & 1 & 1 & 0 & 0\cr
1 & 0 & 1& 0 & 1 & 0 & 1\cr
0 & 1 & 1 & 0 & 1 & 1 & 0\cr
1 & 1 & 0 & 0 & 0 & 1 & 1} \ ,\quad
M\inv={1\ov 4} \pmatrix{-1& 1 & -1 & 1 & 1 & -1& 1\cr
1& -1& -1& 1& -1& 1& 1\cr
1 & 1 &-1& -1 & 1 & 1 & -1\cr
1 & 1 & 1 & 1 & -1 & -1 & -1\cr
-1 & -1 & 1& 1 & 1 & 1 & -1\cr
-1 & 1 & 1 & -1 & -1 & 1 & 1\cr
1 & -1 & 1 & -1 & 1 & -1 & 1} \ .
\ee
The matrix $M$ has in each row four non-zero unit elements
precisely at the column position corresponding to the indices for
which the elements of $\psi_{abcd}$, given in \eqn{g2str2}, are non-zero.

Then our system of differential equations \eqn{om12}
becomes\footnote{We note is passing the useful identities
\ba
&& M_{ac} M_{bc}= 2(\d_{ab}+1)\ ,
\nonumber\\
&&
{1\ov 6} \psi_{abc}^2 M\inv_{ad}M\inv_{be}M\inv_{cf} = -{1\ov 6} \d_{de}\d_{df}
+{1\ov 48}(\d_{de}+\d_{df}+\d_{ef}) -{1\ov 192}\ ,
\ea
where, in the second identity we sum over the repeated indices
$a,b,c$ on the left hand side. There is no sum on the right hand side.
In addition, we note that the system \eqn{aai} was also derived in \cite{Ueno}.
}
\be
{d\Om_a \ov d\tau} = {1\ov 4} \Om \Om_a - \Om_a^2 \ , \quad
 a =1,2,\dots , 7\ ,\qq \Om\equiv
\sum_{b=1}^7 \Om_b \ .
\label{aai}
\ee
This
is a particular case of a more general system which is completely integrable
as we will show.
Namely, let us consider the $N$-dimensional system of
first order non-linear differential equations
\be
{d\Om_a \ov d\tau} = {1\ov \D} \Om \Om_a - \Om_a^2 \ , \quad
 a =1,2,\dots , N\ ,\qq \Om\equiv
\sum_{b=1}^N \Om_b \ ,
\label{suu1}
\ee
where $\D$ is a real constant.
This is also a gradient flow in the sense of \eqn{grom}-\eqn{grv} with
the $\Om_a$'s replacing $\om_a$'s and the summation extended as $a,b=1,2,\dots,
N$. The prepotential, the metric and its inverse are
\ba
W& =&  -{1\ov 6} \sum_{a=1}^N \Om_a^3 + {1\ov 4\D}
 \Om \sum_{a=1}^N \Om_a^2 -{1\ov 12 \D^2} \Om^3\ ,
\nonumber\\
g_{ab}& = & 2 \d_{ab}+ {2\ov 2\D-N}\ ,\qq
g\inv_{ab}=\ha \d_{ab} - {1\ov 4\D} \ ,
\ea
where we exclude of course the case $N=2\D$ since then the
metric $g_{ab}$ has no meaning.

It is remarkable that a system identical to \eqn{suu1}
arose before in the completely different context of constructing domain
wall solutions in gravity coupled to scalars
theories, corresponding to sectors of gauged supergravities in four, five
and seven dimensions \cite{bs1,bs2}.
When lifted to string or M-theory, these
solutions have the interpretation as the gravitational field of continuous
distributions of $p$-branes.\footnote{For work related to these type of
solutions in relation to the Coulomb branch of $\cN=4$ SYM
see also \cite{fgpw2,brand1,cvgu}.}
In this analogy, the r\^ole of the functions
$\Om_a$ is played by exponentials of scalars leaving in the coset
$SL(N,\IR)/SO(N)$ which are used to deform the $SO(N)$ spherical symmetry of
the space transverse to the branes in the ten- or eleven-dimensional
supergravity solutions.\footnote{Specifically, the system \eqn{suu1}
coincides with the system (3.2) of \cite{bs2} if the various
parameters and functions of \cite{bs2} are identified as: $z=-\tau$, $g=1$,
$e^{2\b_i}\to \Om_a f^{1/N-1/\D}$. Then, the solution in eq. (3.6) of
\cite{bs2} also coincides with \eqn{gfeee}.}
Guided by these previous works we find that,
for all values of $N$ and $\D$,
the most general solution to the system \eqn{suu1} is given by
\be
\Om_a = {f^{1/\D}\ov F-b_a}\ , \quad a=1,2,\dots , N\ ,
\qq  f\equiv \prod_{c=1}^N (F-b_c)\ ,
\label{gfeee}
\ee
where the function $F(\tau)$ satisfies the differential equation
\be
\left({dF\ov d\tau}\right)^\D = f\ .
\label{gfe}
\ee
The $b_a$'s are the $N$ constants of integration, which, without
loss of generality, can be ordered as
\be
b_1\ge b_2\ge \cdots \ge b_N\ .
\label{arrra}
\ee
Hence, the entire problem boils down to solving the differential equation
\eqn{gfe}. In general, this is a difficult task, to which we will shortly
turn. We also mention that, a solution to the system \eqn{aai} was also
presented in \cite{Ueno} in terms of an auxiliary function satisfying
a non-linear differential equation.
Presumably it
is equivalent to ours for $N=7$ and
$\D=4$ by appropriate transformations and renaming of variables.

Note that, when all the $b_a$'as are equal, then the $SO(N)$
rotational symmetry in
the $N$-dimensional Euclidean space spanned by the $\Om_a$'s is preserved.
This
symmetry breaks into smaller subgroups when some of the $b_a$'s
differ from each other
and it is completely broken in the generic case when all of them are unequal.
We point out that the corresponding symmetry in the space of the $\om_a$'s
is usually smaller since the transformation \eqn{ttrr} doesn't preserve
rotations.

It is consistent to set to zero in the system \eqn{aai}
some of the $\Om_a$'s and then proceed to solve for the rest, as before.
In such cases, the solution is still given by \eqn{gfeee} and \eqn{gfe}
with the same value for $\D$, but with less moduli parameters $b_a$
corresponding to a smaller value for $N$.
It is possible to recover these particular cases from our general solution
\eqn{gfeee} by a limiting procedure.
To be concrete, let us consider the rescaling (for the generic case with
$N\neq \D+1$)
\be
F\to F (-b_N)^{-{1\ov N-\D-1}}\ , \qq b_a= b_a (-b_N)^{-{1\ov N-\D-1}}\ ,
\quad  a =1,2,\dots , N\!-\!1\ ,
\ee
and then take the limit $b_N\to -\infty$. Then, from
the solution \eqn{gfeee} we obtain that $\Om_N=0$, whereas for $a=1,2,\dots ,
N\!-\!1$ the expressions are the same, but with $N$ replaced by $N-1$.
Also, after the limit, the differential equation \eqn{gfe}
does not contain in its right hand side the factor $(F-b_N)$.
This of course is the solution we would have found if we had
started from the very beginning with $\Om_N=0$.

\section{Examples}
\setcounter{equation}{0}

Let us now focus to the case of interest, the 7-dimensional Euler problem,
and set the parameters to their values $N=7$ and $\D=4$.
We've seen that the problem boils down mathematically to solving the
differential equation \eqn{gfe}.
A more detailed study of \eqn{gfe} necessarily involves techniques
of algebraic geometry.
This task will be undertaken in the following section
which can be read independently.
In this section we will mention some
general features and give a few elementary examples.

The evolution of the function $F(\tau)$
should be such that the $\Om_a$'s remain real which in turn implies that
$f\ge 0$. The function $F(\tau)$ can be bounded by two
consecutive constants among the
$b_a$'s, for instance $b_{2}\le F\le b_1$, or it can be unbounded.
In the latter case we have necessarily that
$F\ge b_1$.
For reasons that will become apparent in the discussion of solitonic
solution in the massive case later in the paper, it is
useful to expose the behaviour of $F(\tau)$ near the end points.
We find that
\be
F= \left(4/3\ov \tau_0-\tau\right)^{4/3}\ ,
\qq {\rm as} \quad \tau\to \tau_0^-\ ,
\label{uuni}
\ee
where $\tau_0$ is a constant of integration.
This is a universal behaviour since it does not
depend on the constants of integration $b_i$.
For $F\to b_1^+$ the behaviour depends crucially on the degree of degeneracy
of $b_1$, which according to the arrangements of parameters in
\eqn{arrra}, is the maximum among the $b_a$'s.
In general, let $b_1=b_2=\cdots = b_n$, so that the degree of
degeneracy of $b_1$ is $n$. Then, for
$n\neq 4$ there is a power law behaviour
\be
F-b_1 = \left((1-n/4) f_0^{1/4} (\tau-\tau_1)\right)^{4\ov 4-n}\ ,
\qq {\rm as} \quad
\tau \to \left\{\begin{array}{lll}
-\infty  & \ \ {\rm if}\  & n=5,6,7\ ,
 \\
\tau_1  & \ \ {\rm if}\ & n=1,2,3  \ ,
\end{array}
\right.
\label{orio}
\ee
where $f_0=\prod_{a=n+1}^7(b_1-b_a)$ and $\tau_1$ is some other constant
related to $\tau_0$ above. The precise relation requires of course the
knowledge of the behaviour of $F(\tau)$ in the entire $\tau$-interval.
For $n=4$ the behaviour is exponential
\be
F-b_1= (\const) e^{f_0^{1/4} \tau}\ , \qq {\rm as} \quad \tau\to -\infty\ .
\label{orio4}
\ee
A similar analysis can be performed for the case of bounded motion,
where $b_2\le F \le b_1$, but will not present it here.

For concreteness we present a few explicit examples.

\no
\underline{\bf Example 1}: In the simplest example all
constants of integration $b_a$
are equal and the $SO(7)$ symmetry in the space of
the $\Om_a$'s remains unbroken. With no loss of generality
we choose them as $b_1=b_2=\cdots =b_7=0$.
Then
\be
F=\left(4/3\ov\tau_0-\tau\right)^{4/3}\
\ee
and
\be
\Om_a=4 \om_a  = {4/3\ov \tau_0-\tau}\ , \qq a=1,2,\dots , 7\ .
\label{iop}
\ee
The result is in agreement with the universal behaviour \eqn{uuni} which
in this case is exact for all values of $\tau$.

\no
\underline{\bf Example 2}: Consider
now the case with $b_1=b_2=b_3=b_4=1$ and $b_5=b_6=b_7=0$.
Then the $SO(7)$ breaks down to the $SO(4)\times SO(3)$ subgroup
and the number of integration constants $b_a$ which equal
the maximum one $b_1$, is $n=4$.
The function $F(\tau)$ is determined by solving \eqn{gfe}.
We have two distinct cases depending on the range of $F$, namely, either
$F\ge 1$ or $0\le F \le 1$.

For $F\ge 1$ the solution is
\be
\ln\left(F^{1/4}+1\ov F^{1/4}-1\right)- 2 \cot\inv\left(F^{1/4}\right)
 = \tau_0-\tau\ ,
\label{h11}
\ee
where $\tau_0$ is an integration constant.
The evolution takes place in the interval $\tau\in (-\infty,\tau_0)$ and
$F(\tau)$ increases monotonically from $1$ to $+\infty$.
We cannot invert \eqn{h11} and obtain $F(\tau)$ explicitly, except
near the end points
\be
F(\tau) = \left\{\begin{array}{lll}
\left(4/3\ov \tau_0-\tau\right)^{4/3} & \ \ {\rm as}\  & \tau \to \tau_0^-\ ,
 \\
1+8 e^{\tau}  & \ \ {\rm as}\ & \tau \to -\infty  \ ,
\end{array}
\right.
\label{limi1}
\ee
which is in agreement with the general expressions \eqn{uuni} and \eqn{orio4}.
Also the $\Om_a$'s in terms of the function $F(\tau)$ are
\be
\Om_1=\Om_2=\Om_3=\Om_4= F^{3/4}\ , \qq \Om_5=\Om_6=\Om_7=(F-1)F^{-1/4}\ .
\ee

For $0\le F\le 1$ the solution is
\be
\ln\left(1+F^{1/4}\ov 1-F^{1/4}\right) +2  \tan\inv\left(F^{1/4}\right)
 = \tau-\tau_0\ ,
\label{h112}
\ee
where, as before, $\tau_0$ is a constant of integration.
The evolution now takes place in the interval $\tau\in (\tau_0,\infty)$ and
$F(\tau)$ increases monotonically from $0$ to $1$.
The behaviour near the end points is
\be
F(\tau) = \left\{\begin{array}{lll}
\left(\tau-\tau_0\ov 4\right)^{4} & \ \ {\rm as}\  & \tau \to \tau_0^+\ ,
 \\
1-8 e^{-\tau}  & \ \ {\rm as}\ & \tau \to +\infty  \ .
\end{array}
\right.
\label{limi2}
\ee
The $\Om_a$'s in terms of the function $F(\tau)$ are
\be
\Om_1=\Om_2=\Om_3=\Om_4=- F^{3/4}\ , \qq \Om_5=\Om_6=\Om_7=(1-F)F^{-1/4}\ .
\ee

\no
\underline{\bf Example 3}: Consider the case with $b_1=b_2=b_3=0$
and $b_4=b_5=b_6=b_7=-1$, where the reality condition forces that $F\ge 0$.
Hence the symmetry subgroup of $SO(7)$ that remains unbroken
is still $SO(4)\times SO(3)$, but now $n=3$. We find that
\be
\Om_1=\Om_2=\Om_3= F^{-1/4} (F+1)\ , \qq \Om_4=\Om_5=\Om_6=\Om_7=F^{-1/4}\
\ee
and
\be
\ln\left(1+\sqrt{2} F^{1/4} + F^{1/2}\ov 1-\sqrt{2} F^{1/4} + F^{1/2}\right)
+ 2 \cot\inv\left(\sqrt{2} F^{1/4}\ov F^{1/2}-1\right)= \sqrt{2}(\tau-\tau_0)
\ .
\ee
The evolution occurs at a finite interval for $\tau$ as $F(\tau)$ grows from
$0$ to $\infty$.

\no
\underline{\bf Example 4}: Consider next the case with
$b_1=b_2=b_3=b_4=b_5=1$ and $b_6=b_7=0$, where necessarily $F\ge 1$.
Now $SO(7)$ is broken to the $SO(5)\times SO(2)$ subgroup and $n=5$.
The solution for $F(\tau)$ is written in terms of a
hypergeometric function ${}_2F_1$ as
\be
F^{-1/2}(F-1)^{-1/4} -{2\ov 3} F^{-3/4} {}_2F_1(3/4,1/4,7/4,1/F) = {1\ov 4}
(\tau_0-\tau)\ .
\label{js2}
\ee
Its behaviour near the end points is
\be
F(\tau) = \left\{\begin{array}{lll}
1+ (-4/\tau)^4 & \ \ {\rm as}\  & \tau \to -\infty\ ,
 \\
\left(4/3\ov \tau_0-\tau\right)^{4/3}  & \ \ {\rm as}\ & \tau \to \tau_0^-  \ ,
\end{array}
\right.
\label{limi3}
\ee
which again
is in agreement with the general expressions \eqn{orio} and \eqn{orio4}.
In between it grows monotonically from $1$ to $+ \infty$.
In terms of $F(\tau)$ the expressions for the $\Om_a$'s are
\be
\Om_1=\Om_2=\Om_3=\Om_4=\Om_5 = (F-1)^{1/4} F^{1/2}\ ,
\qq \Om_6=\Om_7 = (F-1)^{5/4} F^{-1/2}
\ .
\ee

\no
\underline{\bf Example 5}: Finally, consider two cases
where we start from the beginning with $\Om_7=0$.
As we have already explained the general solution is given by
\eqn{gfeee} and \eqn{gfe} with $N=6$ and has six moduli parameters.
First we choose them such that
$b_1=b_2=b_3=b_4=1$ and $b_5=b_6=0$, so that the subgroup of the maximum
symmetry group $SO(6)$ which is preserved is $SO(4)\times SO(2)$ and
$n=4$. We have two cases depending on whether $F\ge 1$ or $0\leq F\leq 1$.
In the former case we find the explicit solution
\be
F(\tau)= \coth^2{\tau\ov 2}\ ,\qq  -\infty< \tau \le 0 ,
\ee
where we have absorbed the integration constant into a redefinition of $\tau$.
From that we compute
\be
\Om_1=\Om_2=\Om_3=\Om_4=F^{1/2}=-\coth{\tau\ov 2}\ ,\qq \Om_5=\Om_6=
(F-1) F^{-1/2}=-{2\ov \sinh\tau}\ .
\label{hjgw}
\ee
In the case with $0\leq F\leq 1$ we find instead that
\be
F(\tau)= \tanh^2{\tau\ov 2}\ ,\qq  0\le \tau < \infty\
\ee
and
\be
\Om_1=\Om_2=\Om_3=\Om_4=-F^{1/2}=-\tanh{\tau\ov 2}\ ,\qq \Om_5=\Om_6=
(1-F) F^{-1/2}={2\ov \sinh\tau}\ .
\label{hjg2}
\ee
Note that this solution can also
be obtained by analytically continue $\tau\to \tau + i\pi $ in \eqn{hjgw}.

\no
\underline{\bf Example 6}: Finally, we consider a generalization of the
previous example such that the preserved symmetry group is enhanced
to $SO(2)\times SO(2)\times SO(2)$. Namely, let $b_1=b_2$, $b_3=b_4$ and
$b_5=b_6$. This is the case of the 3-dimensional
Euler top since using \eqn{ttrr} we find
that $\om_1=\om_2=\om_6=\om_7 = 0$ and that the remaining $\om_4$, $\om_5$ and
$\om_6$ obey the standard 3-dimensional Euler top
equations in \eqn{eell} (with $\om_i$
replaced by $\om_{i+3}$).
In that case we know of course that the solution
is given in terms of elliptic Jacobi functions with the help of
the two independent constants of motion
\be
I_1=\om_3^2 -\om_4^2\ ,\qq I_2=\om_4^2 -\om_5^2\ .
\label{i1i2}
\ee
This will be verified in our framework.
Indeed, in the case at hand, \eqn{gfe} reduces to the Weierstrass
differential equation with solution
\be
F(\tau)=\wp(\tau/2) + {1\ov 3} (b_1+b_3+b_5)\ ,
\label{st3D}
\ee
where $\wp(\tau/2)$ is the Weierstrass function. This is a doubly
periodic function in the argument $\tau/2$ and
the two half-periods are given by\footnote{The half-periods $\om_1$
and $\om_2$ below should not be confused with the $\om_a(\tau)$'s
that parametrize the gauge fields.}
\be
{\rm half\!-\!periods}:\qq
\om_1 = {K(k)\ov \sqrt{e_1-e_3}}\ ,\qq \om_2 = {i K(k')\ov \sqrt{e_1-e_3}}
\ ,
\label{wjd1}
\ee
where $K$ is the complete elliptic integral of the first kind with modulus
$k$ and complementary modulus $k'$ given by
\be
k^2 = {e_2-e_3\ov e_1-e_3}\  ,
\qq
k'^2=1-k^2 = {e_1-e_2\ov e_1-e_3}\ .
\label{wekdfj12}
\ee
Here $e_1$, $e_2$ and $e_3$ are the values of the  Weierstrass function
at the half-periods, i.e. $\wp(\om_1)=e_1$, $\wp(\om_2)=e_3$ and
$\wp(\om_1+\om_2)=e_2$, which
are expressed in our case in terms of the parameters $b_1$, $b_3$ and $b_5$
as
\be
e_1 = b_1 -{1\ov 3} (b_1+b_3+b_5)\ .
\label{jefh2}
\ee
The expressions for $\e_2$ and $e_3$ are given by the above formulae
after interchanging the r\^oles of $b_1$ with $b_3$ and $b_5$,
respectively.

Then we find that
\ba
&& \Om_1=\Om_2 = \sqrt{(F-b_3) (F-b_5)\ov F-b_1} =
(b_1-b_5)^{1/2}\ {{\rm dn}u \ov {\rm cn}u\ {\rm sn}u}\ ,
\nonumber\\
&&\Om_3=\Om_4 =  \sqrt{(F-b_1) (F-b_5)\ov F-b_3} =
(b_1-b_5)^{1/2}\ {{\rm cn}u \ov {\rm dn}u\ {\rm sn} u}\ ,
\nonumber\\
&&\Om_5=\Om_6 =  \sqrt{(F-b_1) (F-b_3)\ov F-b_5} =
(b_1-b_5)^{1/2}\ {{\rm cn}u\ {\rm dn} u\ov {\rm sn} u}\ ,
\nonumber\\
&& u\equiv -\ha (b_1-b_5)^{1/2}\ \tau + u_0 \ ,\qq \tau\le 0\ ,
\ea
with $u_0$ being a constant of integration and
where we have used well known properties of the Weierstrass function to
express the final result in terms of the Jacobi functions
${\rm sn}u$, ${\rm cn}u$ and ${\rm dn} u$. Also, using \eqn{ttrr} we find that
\be
\om_3=\ha(\Om_1+\Om_5-\Om_3)\ ,\qq \om_4=\ha (\Om_1+\Om_3-\Om_5)\ ,\qq
\om_5=\ha (\Om_3+\Om_5-\Om_1)\ ,
\ee
and zero for the rest. It is easily verified, using properties of the
Jacobi functions, that the constants of motion \eqn{i1i2}
are $I_1=b_5-b_3$ and $I_2=b_1-b_5$.

In the particular case with $b_1=b_2=b_3=b_4=1$ and $b_5=b_6=0$
the complementary modulus $k'\to 0$ and
we may use the approximate expressions for the Jacobi functions
${\rm sn}u\simeq \tanh u$ and ${\rm dn}u\simeq
{\rm cn}u\simeq 1/\cosh u$. Choosing the constant $u_0$ to be real and then
absorbing it into a redefinition of $\tau$ we
recover \eqn{hjgw} as we should.
In the case with
$b_1=b_2=1$ and $b_3=b_4=b_5=b_6=0$ the modulus $k\to 0$ and we may use
instead the approximate expressions
${\rm sn}u\simeq \sin u$, ${\rm cn}u\simeq \cos u$ and ${\rm dn}u\simeq 1$.
Then we find that
\be
\Om_1=\Om_2=-{2\ov \sin\tau}\ ,\qq \Om_3=\Om_4=\Om_5=\Om_6=-\cot {\tau\ov 2}\ ,
\qq -{\pi \ov 2}\le \tau \le 0 \ ,
\ee
which is the trigonometric counterpart of \eqn{hjgw}. The solution \eqn{hjg2}
can also be obtained in this limit if we choose the constant of integration
purely imaginary, namely $u_0=i \pi/2$.
This essentially amounts to the analytic continuation
in $\tau$,  we noted before, that relates \eqn{hjgw} and \eqn{hjg2}.

\section{Algebraic curve classification}
\setcounter{equation}{0}

In this section we systematically study the differential equation \eqn{gfe}
using classical techniques from algebraic geometry.
We will see that there are cases where this approach is not
only mathematically elegant, but also serves as a practical tool in
explicit computations.
We basically follow the
discussion in \cite{bs1}, where such an equation arose in the construction
of domain wall solutions in gauged supergravity.
An approach, though not quite as systematic, based on algebraic curves was
also followed in \cite{Ueno}.

The non-linear differential equation \eqn{gfe}
can be viewed, when the parameter $\tau$ and the unknown function $F(\tau)$
are extended to the complex domain, as a Christoffel--Schwarz
transformation from a closed polygon in the $\tau$-plane
onto the upper-half $F$-plane.
In this
case the perimeter of the polygon is mapped to the real $F$-axis,
whereas its vertices are mapped to points parametrized by the
moduli $b_i$.
Letting
\be
x=F(\tau)\ ,\qq  y={d F(\tau)\ov d\tau}\ ,
\label{xtyt}
\ee
 we arrive at the algebraic curve in $C^2$
\be
y^4 = \prod_{a=1}^7 (x -b_a)\ .
\label{alcu1}
\ee
Given this algebraic curve we are faced with
the problem of multi-valuedness since the corresponding
Riemann surface is pictured geometrically by gluing four sheets together
along their branch cuts. This is resolved in a standard way by uniformizing
the algebraic curve in terms of the so called
uniformizing complex parameter, say $u$.
As a first step one constructs birational invertible
transformations $x(v,w)$, $y(v,w)$ such that the
algebraic curve \eqn{alcu1} assumes in terms of the new set of
variables ($v$,$w$) the canonical standard form according to the genus $g$
of the associated to the albegraic curve Riemman surface.
For instance, according to algebraic geometry every $g=0$ Riemman
surface can be brought into the form $u=v$. Then we may use as uniformizing
parameter $u=v=w$. Similarly, every $g=1$ Riemman
surface can be brought into the Weierstrass forms $w^2=4 v^3 -g_2 v-g_3$
or $w^2= 4(v-e_1)(v-e_2)(v-e_3)$.
Then the uniformization problem is solved as $v=\wp(u)$ and $w=\wp'(u)$,
where $\wp(u)$ is the Weierstrass function and $u$ denotes again the
uniformizing parameter.
In general, every Riemann surface with $g\le 2$ can
brought into the hyperelliptic form which is still tractable.
However, as the genus of the Riemann surface increases the uniformization
problem becomes increasingly more difficult to solve.
In order to relate the uniformizing parameter $u$ to $\tau$, we form,
after solving the uniformization problem, the functions
$x = x(u)$ and $y = y(u)$ and restrict the domain of values for $u$, so that
$x=F(\tau)=x(u)$ is a real function.
Then we may obtain $u(\tau)$ by inverting the solution of the
differential equation
\be
{d\tau \over du} = {1 \over y(u)}{dx(u) \over du} \ ,
\label{bouco}
\ee
which is derived using \eqn{xtyt}.

The genus of the curve can be easily determined
via the Riemann--Hurwitz relation. Recall that for any
curve of the form
\be
y^m = (x-\l_1)^{\a_1}(x-\l_2)^{\a_2} \dots (x-\l_n)^{\a_n}\ ,
\label{genc}
\ee
with integers $m$ and $\a_i$ having no common factors,
and all $\l_i$'s being unequal,
the genus $g$ can be found by first writing the ratios
\be
{\a_1 \over m} = {d_1 \over c_1} ~, ~~ \cdots ~~ , {\a_n \over m} =
{d_n \over c_n} ~; ~~~~~~ {\a_1 + \cdots + \a_n \over m} = {d_0 \over c_0}
\ee
in terms of relatively prime numbers and then using the relation
\be
g = 1 - m +{m \over 2} \sum_{i=0}^n \left(1-{1 \over c_i}\right) .
\ee
According to this, the genus of our surface turns out to be $g=9$
when all $b_i$ are unequal,
and so it is difficult to determine explicitly the
solution in the general case. However, when some the
parameters among the $b_i$'s are equal the genus becomes smaller
since that corresponds to degenerating the surface along certain cycles, thus
reducing its genus. Similarly, the symmetry group of the solution
\eqn{gfeee} gets enhanced into larger subgroups of $SO(7)$.
In such cases the problem becomes more tractable. The complete list of the
various algebraic curves and the genus of the associated Riemann
surfaces is given by the table below. It is identical to the table 4
of \cite{bs3} corresponding to distributions of D2-branes.

\begin{center}
\begin{tabular}{c|l|l}
{\hskip-3pt \em Genus} & {\em Irreducible Curve} & {\em Isometry Group}\\
\hline
 9 & $y^4=(x-b_1)(x-b_2) \cdots (x-b_6)(x-b_7)$ & None \\
\hline
 7 & $y^4=(x-b_1)(x-b_2) \cdots (x-b_5)(x-b_6)^2$ & $SO(2)$ \\
\hline
 6 & $y^4=(x-b_1)(x-b_2) \cdots (x-b_4)(x-b_5)^3$ & $SO(3)$ \\
\hline
 5 & $y^4=(x-b_1) \cdots (x-b_3)(x-b_4)^2(x-b_5)^2$
& $SO(2) \times SO(2)$ \\
\hline
 4 & $y^4=(x-b_1)(x-b_2)(x-b_3)^2(x-b_4)^3$
& $SO(2) \times SO(3)$ \\
\hline
 3 & $y^4=(x-b_1) \cdots (x-b_3)(x-b_4)^4$ & $SO(4)$ \\
    & $y^4=(x-b_1)(x-b_2)(x-b_3)^5$ & $SO(5)$ \\
    & $y^4=(x-b_1)(x-b_2)^3(x-b_3)^3$
& $SO(3) \times SO(3)$ \\
    & $y^4=(x-b_1)(x-b_2)^2(x-b_3)^2(x-b_4)^2$
& $SO(2)^3$ \hskip .5cm \\
\hline
 2 & $y^4 = (x-b_1)^2(x-b_2)^2(x-b_3)^3$
& $SO(2) \times SO(2) \times SO(3)$ \\
\hline
 1 & $y^4=(x-b_1)(x-b_2)^6$ & $SO(6)$ \hskip .5cm \\
    & $y^4= (x-b_1)(x-b_2)^2(x-b_3)^4 $
& $SO(2) \times SO(4)$ \hskip .5cm \\
& $y^4=(x-b_1)^2 (x-b_2)^5$ &$SO(2) \times SO(5)$  \hskip .5cm \\
\hline
 0     & $y^4=(x-b_1)^3 (x-b_2)^4$ &$SO(3) \times SO(4)$  \hskip .5cm \\
       &  $y^4 = (x-b_1)^7$ & $SO(7)$  (maximal) \\
\hline
\end{tabular}
\end{center}
\begin{center}
Table 1: Curves, their genus and symmetry groups for $N=7$ moduli.
\end{center}
According to this table,
the first three examples we considered explicitly in section 3
correspond to $g=0$ Riemann surfaces, whereas the fourth one to a
surface with $g=1$. This is also reflected in the fact that in the $g=0$
cases the relation between $F$ and $\tau$ is via elementary functions,
whereas in the $g=1$ case the hypergeometric special function is involved.

Similarly, we may consider the case of $N=6$ moduli parameters. Then the
generic case when all the moduli parameters are unequal
corresponds to a $g=7$ Riemann surface. The resulting table 2
below arises also
in studies of domain wall solutions in five-dimensional gauged supergravity
that correspond to continuous distributions of D3-branes in type-IIB string
theory \cite{bs1}.

\begin{center}
\begin{tabular}{c|l|l}
{\em Genus} & {\em Irreducible Curve} & {\em Isometry Group}\\
\hline
 7 & $y^4=(x-b_1)(x-b_2) \cdots (x-b_5)(x-b_6)$ & None \\
\hline
 5 & $y^4=(x-b_1)(x-b_2)(x-b_3)(x-b_4)(x-b_5)^2$ & $SO(2)$ \\
\hline
 4 & $y^4=(x-b_1)(x-b_2)(x-b_3)(x-b_4)^3$ & $SO(3)$ \\
\hline
 3 & $y^4=(x-b_1)(x-b_2)(x-b_3)^2(x-b_4)^2$
& $SO(2) \times SO(2)$ \\
\hline
 2 & $y^4=(x-b_1)(x-b_2)^2(x-b_3)^3$
& $SO(2) \times SO(3)$ \\
\hline
 1 & $y^4=(x-b_1)(x-b_2)(x-b_3)^4$ & $SO(4)$ \\
    & $y^4=(x-b_1)(x-b_2)^5$ & $SO(5)$ \\
    & $y^4=(x-b_1)^3(x-b_2)^3$
& $SO(3) \times SO(3)$ \\
    & $y^2=(x-b_1)(x-b_2)(x-b_3)$
& $SO(2)^3$ \\
\hline
 0 & $y^2 = (x-b_1)(x-b_2)^2$
& $SO(2) \times SO(4)$ \\
    & $y^2=(x-b_1)^3$ & $SO(6)$ (maximal) \\
\hline
\end{tabular}
\end{center}
\begin{center}
Table 2: Curves, their genus and symmetry groups for $N=6$ moduli.
\end{center}
According to this table, the fifth example we gave in section 3
corresponds to a $g=0$ surface,
whereas the sixth one to a $g=1$ surface.
In these cases the uniformization procedure that we have advocated and
described in this section is not necessary as the solution can be readily
found by elementary methods. However, finding solutions to the other cases
requires that the uniformization program, which turns out to be non-trivial,
is carried out completely. As an example let's consider the other symmetric
case with $g=1$ with algebraic curve $y^4=(x-1)^3x^3$, where for convenience
we have chosen $b_1=1$ and $b_2=0$ (this can always be done with
appropriate rescalings and shifts).
We find that the birational transformation
\be
x=1-{1\ov v} (v+1/4)^2\ ,\qq y={w^3\ov 8 v^3}\ ,
\label{biio}
\ee
brings the curve into the form $w^2=4 v^3 -v/4$ which is of the standard
Weierstrass form with
\ba
&&g_3=0\ ,\qq g_2={1\ov 4}\ ,\qq e_1={1\ov 4}\ , \qq e_2=0\ ,
\qq e_3=-{1\ov 4}\ ,
\nonumber\\
&& k=k'={1\ov \sqrt{2}}\ , \qq \om_1=-i \om_2 = {\G(1/4)^2\ov 2 \sqrt{2\pi}}\ ,
\label{bio1}
\ea
where we have included the values for the modulus, its equal complementary
modulus and the two half periods. We have then that $x=\wp(u)$ and $y=\wp'(u)$,
where $u$ is determined as a function of $\tau$ after solving \eqn{bouco}.
In this case it turns out that we get the simple relation $u=-\tau/2$.
Hence, we found that
\be
F(\tau)=1-{1\ov \wp(\tau/2)}(\wp(\tau/2)+1/4)^2=
1-{2\ov {\rm sn}^2\left(\tau\ov 2 \sqrt{2}\right)
{\rm dn}^2\left(\tau\ov 2 \sqrt{2}\right)}\ ,
\ee
where we have used the evenness of the Weierstrass function and,
in the last step, its relation to the Jacobi functions.
For completeness we also mention the inverse of the transformation
\eqn{biio}
\be
v={1\ov 4} {y^2-(x-1)x^2\ov y^2+(x-1)x^2}\ , \qq
w={(x-1)x\ov 2 y}\  {y^2-(x-1)x^2\ov y^2+(x-1)x^2}\ .
\ee
The reader is referred to section 4 of \cite{bs1}
for the other low genus cases $g=0,1,2$ of table 2, where the uniformization
procedure has been done explicitly.

Similarly, for the cases with $N=5$ moduli parameters we have formed table 3.
In the same spirit as before, this is identical to table 3
of \cite{bs2}, that arose in the construction of
domain-wall solutions of seven-dimensional gauged supergravity
corresponding to continuous distributions of M5-branes in M-theory.

\begin{center}
\begin{tabular}{c|l|l}
{\em Genus} & {\em Irreducible Curve} & {\em Isometry Group}\\
\hline
 6 & $y^4=(x-b_1)(x-b_2)(x-b_3)(x-b_4)(x-b_5)$ & None \\
\hline
 4 & $y^4=(x-b_1)(x-b_2)(x-b_3)(x-b_4)^2$ & $SO(2)$ \\
\hline
 3 & $y^4=(x-b_1)(x-b_2)(x-b_3)^3$ & $SO(3)$ \\
\hline
 2 & $y^4=(x-b_1)(x-b_2)^2(x-b_3)^2$ & $SO(2)^2$ \\
\hline
 1 & $y^4=(x-b_1)^2(x-b_2)^3$ & $SO(2) \times SO(3)$  \\
\hline
 0 & $y^4=(x-b_1)(x-b_2)^4$ & $SO(4)$ \\
    & $y^4=(x-b_1)^5$ & $SO(5)$  (maximal) \\ \hline
\end{tabular}
\end{center}
\begin{center}
Table 3: Curves, their genus and symmetry groups for $N=5$ moduli.
\end{center}
As an example, consider the $g=1$ curve $y^4=(x-1)^2 x^3$.
The transformation
\be
x=1-{w^2\ov 4 v^3} \ , \qq y={w\ov v}\left(1-{w^2\ov 4 v^3}\right)\ ,
\ee
with inverse
\be
 v=- {x(x-1)\ov 4 y^2}\ ,\qq w=-{x-1\ov 4 y} \ ,
\ee
brings the curve into the same standard form $w^2=4 v^3 -v/4$ as in the
previous example in this section. Hence $v=\wp(u)$ and  $y=\wp'(u)$,
whereas the various parameters are still given
by \eqn{bio1}. It also turns out that $u=-\tau/2$ as before.
Finally, the result for $F(\tau)$ is
\be
F(\tau)= {1\ov 16 \wp(\tau/2)^2}= {{\rm sn}^4 \left(\tau\ov 2\sqrt{2}\right)
\ov 4 {\rm dn}^4 \left(\tau\ov 2 \sqrt{2}\right)} \ .
\ee
For other examples with $g=0,1$,
where the uniformization program has been carried out explicitly,
we refer the reader to section 7 of \cite{bs2}.

Similarly, the case with $N=4$ moduli parameters leads to table 4 below and
the generic Riemann surface has $g=3$. It coincides with table 5 of
\cite{bs3} that arose in the construction of supergravity solutions
corresponding to continuous distributions of NS5- and D5-branes
in string theory.

\begin{center}
\begin{tabular}{c|l|l}
{\em Genus} & {\em Irreducible Curve} & {\em Isometry Group}\\
\hline
 3 & $y^4=(x-b_1)(x-b_2)(x-b_3)(x-b_4)$ & None \\
\hline
 1 & $y^4=(x-b_1)(x-b_2)(x-b_3)^2$ & $SO(2)$ \\
\hline
 0 & $y^4=(x-b_1)(x-b_2)^3$ & $SO(3)$ \\
    & $y^2=(x-b_1)(x-b_2)$ & $SO(2)\times SO(2)$   \\
    & $y=(x-b_1)$ & $SO(4)$ (maximal) \\
\hline
\end{tabular}
\end{center}
\begin{center}
Table 4: Curves, their genus and symmetry groups for $N=4$ moduli.
\end{center}

Similarly, the case of $N=3$ moduli parameters leads to table 5 below and
the generic Riemann surface has again $g=3$. It coincides with table 6 of
\cite{bs3} that arose in the construction of solutions representing
D6-branes in string theory.

\begin{center}
\begin{tabular}{c|l|l}
{\em Genus} & {\em Irreducible Curve} & {\em Isometry Group}\\
\hline
 3 & $y^4=(x-b_1)(x-b_2)(x-b_3)$ & None \\
\hline
 1 & $y^4=(x-b_1)(x-b_2)^2$ & $SO(2)$   \\
\hline
 0 & $y^4=(x-b_1)^3$ & $SO(3)$ (maximal) \\
\hline
\end{tabular}
\end{center}
\begin{center}
Table 5: Curves, their genus and symmetry groups for $N=3$ moduli.
\end{center}

Finally, we include in table 6 below the case
of $N=2$ moduli parameters that generically
corresponds to a $g=1$ Riemann surface.

\bigskip
\begin{center}
\begin{tabular}{c|l|l}
{\em Genus} & {\em Irreducible Curve} & {\em Isometry Group}\\
\hline
 1 & $y^4=(x-b_1)(x-b_2)$ & None \\
\hline
  0 & $y^2=(x-b_1)$ & $SO(2)$ (maximal) \\
\hline
\end{tabular}
\end{center}
\begin{center}
Table 6: Curves, their genus and symmetry groups for $N=2$ moduli.
\end{center}
In this case the transformation
\be
x=-{1\ov v} (v-1/4)^2 \ , \qq y={w\ov 2 v} \ ,
\ee
with inverse
\be
v={1\ov 4}(2 y^2-2 x+1)\ , \qq w=\ha y(2 y^2-2 x+1)\ ,
\ee
brings the $g=1$
curve $y^4=(x-1)x $ into the Weierstrass form $w^2=4 v^2-v/4$ and
hence $v=\wp(u)$, $w=\wp'(u)$ solve the uniformization problem with parameters
given by \eqn{bio1}.
The differential equation \eqn{bouco} takes in this case the form
$d\tau/du= -\ha \wp'(u)^2/\wp(u)^2$. Integrating leads to an
expression for $\tau(u)$ in
terms of elliptic integrals.  However, inverting this expression and
obtaining the function $u(\tau)$ explicitly is not possible.

\section{The massive case and solitons}
\setcounter{equation}{0}

Let us modify the first order equations \eqn{om122} by a mass term as
\be
{dA_a\ov d\tau} = \ha \psi_{abc}[A_b,A_c] + m_{ab} A_b \ ,
\ee
where $m_{ab}=m_{ba}$ is a constant matrix (this mass term is similar
to the symmetry breaking term in \cite{zachos}
for membranes embedded in 8-dimensions).
Writing as before that $A_a = \psi_a \om_a$ (no num)
we obtain the system
\be
{d\om_a\ov d\tau} = \ha \psi_{abc}^2 \om_b \om_c + m_{ab} \om_b \ ,
\label{11or}
\ee
which can be derived from the prepotential
\be
W= {1\ov 6} \psi^2_{abc} \om_a \om_b \om_c + \ha m_{ab} \om_a\om_b\ .
\ee
The corresponding potential can be computed using \eqn{grv}.

The solution of \eqn{11or} in the
case of a matrix proportional to the identity, i.e.
$m_{ab}= m \d_{ab}$, can be obtained from that
of the corresponding massless case \eqn{om12}. Indeed,
after performing the change of variables
\be
\om_a =e^{m \tau} \bar \om_a  \ ,\qq \eta={e^{m\tau}\ov m}\ ,
\label{klp}
\ee
we observe that the $\bar \om_a$'s obey
\be
{d\bar \om_a\ov d\eta}=
\ha \psi^2_{abc} \bar\om_b \bar\om_c\ ,\qq a,b,c=1,2,\dots ,7\ ,
\ee
which is just the system \eqn{om12} that we have solved.
For more general mass matrices we know of no such similar transformation.
We remark also that \eqn{klp} is similar to the transformation
employed in \cite{Bachas} in order to convert the
BPS condition for dimensionally reduced $\cN=1$ YM to the usual Nahm system
\eqn{nahmm}.

Adding a mass term has a physical motivation. A diagonal matrix
$m_{ab}=m \d_{ab}$ arises naturally when the self-duality conditions are
defined on an
8-dimensional Euclidean $AdS_8$ space with metric
$ds^2=d\tau^2+ e^{-2m\tau} \sum_{a=1}^7 dx_a^2$, corresponding to a
cosmological constant $\L\sim -m^2$.
We remark that, unlike the 4-dimensional case \cite{EgH},
the energy momentum tensor is not
zero when the gauge field strength satisfies the self-duality conditions in
8-dimensions in some gravitational background of Euclidean signature. Hence
the Einstein's equations in the presence of a negative cosmological
constant are not satisfied with an $AdS_8$ metric.
Therefore, in the statement above, $AdS_8$ is considered as a
{\it fixed} background.

\subsection{Solitons}

In the massless case there are no isolated degenerate vacua corresponding to
minima of the potential. However, such vacua
develop when a mass term is turned on and in
these cases we expect that there are solitonic solutions to the equations of
motion which interpolate between them.
The size of these solitons will of course be dictated by the interplay
between the various mass parameters.
To be concrete,
we will consider the most symmetric
case of a diagonal mass matrix $m_{ab}=m \d_{ab}$, but a similar discussion
can be made for other choices as well.
Then, the
prepotential has seven critical points labeled by the
integer $n=0,1,2,3,5,6,7$. They are found by solving the system of seven
algebraic equations obtained by setting the right hand side of \eqn{aai} to
zero. Without loss of generality these critical points can be arranged to be
\ba
&& \Om_a^{(n)}={4\ov 4-n}m \ ,\qq a=1,2,\dots , n\ ,
\nonumber\\
&& \Om_{n+1}^{(n)}=\Om_{n+2}^{(n)}= \cdots = \Om_7^{(n)} =0\ ,
\label{vaa2}
\ea
up to a renaming of the index $a$.
Hence in the $n$th vacuum the maximal symmetry $SO(7)$ is broken to
$SO(n)\times SO(7-n)$.
In terms of the $\om_a$'s, using \eqn{ttrr},
we find that these critical points
are
\ba
n=0  & :& \qq \vec\om^{(0)} =( 0,0,0,0,0,0,0) \ , \phantom{xxxxxxxxxxx}SO(7)\ ,
\nonumber\\
n=1  & :& \qq \vec\om^{(1)} ={m\ov 3}(-1,1,1,1,-1,-1,1)\ ,
\phantom{xxxxx}SO(4)\times SO(3)\ ,
\nonumber\\
n=2  & :& \qq \vec\om^{(2)} =m(0,0,1,1,-1,0,0)\ ,
\phantom{xxxxxxxx} SO(4) \times SO(2)\ ,
\nonumber\\
n=3  & :& \qq \vec\om^{(3)} =m(- 1,-1,1,3,-1,1,1)\ ,
\phantom{xxxxx}SO(3)\times SO(3)\ ,
\label{vaacc}\\
n=5  & :& \qq \vec\om^{(5)} =m(-1,1,-1,-3,-1,1,-1)\ ,
\phantom{xx} SO(4) \times SO(2)\ ,
\nonumber\\
n=6  & :& \qq \vec\om^{(6)} =m(0,0,-1,-1,-1,0,0)\ ,
\phantom{xxxxx} SO(4)\times SO(3)\ ,
\nonumber\\
n=7  & :& \qq \vec\om^{(7)} =-{m\ov 3}( 1,1,1,1,1,1,1)\ ,
\phantom{xxxxxxxx} SO(7)\ .
\nonumber
\ea
In the last column above we indicated the subgroup of $SO(7)$ which is
preserved by the corresponding vacuum.
Notice that, the symmetry group is the same for the $n$th and
the $(7\!-\!n)$th vacua. Also, the symmetry of a vacuum in the space
of the $\om_a$'s is different than that in the space of $\Om_a$'s
we stated above.
The reason, as we have mentioned, is that the transformation \eqn{ttrr}
breaks part of this symmetry.
One can expand perturbatively around each vacuum separately and derive the
corresponding mass matrix. The matrix elements for the $n$th vacuum
are given by
\be
M^{(n)}_{ab} = m\d_{ab} + \psi_{abc}^2 \om^{(n)}_c \ ,
\qq a,b,c=1,2,\dots , 7\ .
\ee
These mass matrices can be diagonalized in each case separately. Their
eigenvalues and their degeneracies are given by
\ba
n=0 &:& \qq \phantom{x}m \ (7\!-\!{\rm fold})\ ,
\nonumber\\
n=1,7 & : & \qq -m,\ {4\ov 3} m\ (6\!-\!{\rm fold})\ ,
\nonumber\\
n=2,6 & : & \qq -m ,\ -2 m , \ 2m\ (5\!-\!{\rm fold})\ ,
\label{massp}\\
n=3,5 & : & \qq -m,\ -4 m\ (2\!-\!{\rm fold}),\ 4m\ (4\!-\!{\rm fold})\ .
\nonumber
\ea
Note that, the mass spectra around vacua having the same symmetry in
\eqn{vaacc} (for instance, for $n=1$ and $n=6$), are {\it not} identical.
There exists instead an identity between the
mass spectrum of the $n$th and the $(8\!-\!n)$th vacuum for $n=1,2,3,5,6,7$.

It can be easily seen that the on-shell action for a solitonic solution
interpolating between the $n$th and the $k$th vacua
above, as $\tau$ goes from $-\infty$ to $\infty$, is finite and given in terms
of the values of the prepotential $W$ at $\tau=\pm \infty$ as
\be
S_{n\to k}=
-W_k + W_n \ ,\qq W_n= W(\tau=-\infty) =-{n(8-n)\ov 3 (n-4)^2} \ m^3\ .
\ee
The expression for $W_k=W(\tau=\infty)$ is similar to the one above
for $W_n$ with $n$ replaced by $k$. Note also the symmetry $W_n=W_{8-n}$ for
$n=1,2,3,5,6,7$, which is similar to the symmetry we noted for the perturbative
mass spectra in \eqn{massp}.

\subsection{Examples}

The simplest solitonic solution is that relating the vacuum
with $\om_a=0\ , a=1,2,\dots , 7$ which is labeled by the integer
$n=0$ in \eqn{vaacc}, with any other vacuum labeled with $n=1,2,3,5,6,7$.
Let us consider for the system \eqn{11or} any of the following 6 truncations
\be
\vec \om(\tau) =  (x(\tau)+\ha) \vec \om^{(n)} \ , \qq n=1,2,3,5,6,7\ ,
\label{truun}
\ee
where $x(\tau)$ is the single function
whose evolution we would like to determine.
It is easy to verify that all $7$ equations in \eqn{11or} give rise to the
same equation for $x(\tau)$ and therefore the above truncations are
consistent. We find that
\be
{d x\ov d\tau}= m (1/4-x^2) \qq \Longrightarrow \qq x(\tau)=
\ha \tanh\left(m(\tau-\tau_0)\ov 2\right)\ .
\label{sooll}
\ee
This is the usual kink solution centered at $\tau=\tau_0$ and
 interpolating between the two vacua
at $x(-\infty)=- \ha$ (equivalently $\vec \om^{(0)}=\vec 0$) and
$x(+\infty)=+\ha$ (equivalently $\vec \om^{(n)}=\vec 0$ with $n$ assuming
one of the
values $n=1,2,3,5,6,7$). The equation in \eqn{sooll} follows from the
prepotential $W=m(x/4 -x^3/3)$ and therefore the
Lagnangian for the theory is (we discard a numerical factor that depends on
$n$ times $m^2$)
\be
\cL= -\ha \dot x^2 - {m^2\ov 2} (x^2-1/4)^2 \ ,
\label{jghd}
\ee
where the familiar ``mexican hat'' potential is computed using
\eqn{grv}. The equation of motion for this Lagrangian
is well known to admit the soliton solution in \eqn{sooll}.
The novelty of our approach is that we found them by solving a first
order equation instead of the second order one that follows from the
Lagrangian \eqn{jghd}.
We also note that, the kink solution in \eqn{sooll} also arises by starting
from the solution \eqn{iop}, i.e. $\bar \om_a=-{1\ov 3\eta}$ and
then using \eqn{klp}.

Another example follows from the consistent truncation to two
fields $x(\tau)$ and $y(\tau)$ as
\be
\Om_1=\Om_2=\cdots =\Om_5 ={4\ov \sqrt{15}} x + {2\ov \sqrt{3}} y\ ,
\qq \Om_6 =\Om_7 = \sqrt{3} y\ ,
\label{jh23}
\ee
where the choice of the arithmetic factors is such that the two fields are
canonically normalized. Indeed, the Lagrangian for the theory takes the
form
\ba
\cL & =&  -\ha(\dot x^2 + \dot y^2) - V\ ,
\nonumber\\
V & =&  {1\ov 6} y^2 (\sqrt{3} m + \sqrt{5} x + y)^2
+{1\ov 120}(2 \sqrt{15} m x + 2 x^2 + 5 y^2)^2\ ,
\ea
where the potential has been computed using \eqn{grv} with prepotential
\be
W= {\sqrt{3}\ov 9} y^3 + {\sqrt{15}\ov 90} (2 x^3 + 15 x  y^2) + {m\ov 2}
(x^2 + y^2)\ .
\ee
Our anzatz \eqn{jh23}, implies that this
prepotential should have as critical points those corresponding
to the cases with $n=0,2,5,7$ in \eqn{vaa2}.
Indeed, we find
\ba
n=0 & : & \qq x=0\ , \phantom{xxxxxxxxx} y=0\ ,
\nonumber\\
n=2 & :& \qq x=-\sqrt{5\ov 3}\ m \ ,  \phantom{xxx}
y ={ 2\ov \sqrt{3}}\ m \ ,
\nonumber\\
n=5 & : & \qq x=-\sqrt{15}\ m \ , \qq y = 0 \ ,\phantom{xxx}
\label{crrq}\\
n=7 & : & \qq x=-{1 \ov 3} \sqrt{5\ov 3}\ m \ , \phantom{xx}
y = -{4\ov 3 \sqrt{3}}
\ m \ .
\nonumber
\ea
Finding solitonic solutions is easy using the replacement \eqn{klp}.
For instance, the solitonic solution
connecting the vacua labeled by $n=0$ and $n=5$ is
\ba
&&\Om_1=\cdots =\Om_5=e^{m\tau}\left(F(\eta)-1\right)^{1/4} F(\eta)^{1/2}\ ,
\qq \Om_6=\Om_7 = e^{m\tau}{\left(F(\eta)-1\right)^{5/4}\ov F(\eta)^{1/2}}\ ,
\nonumber\\
&&
\eta=e^{m\tau}/m\ , \qq m<0\ ,
\ea
where the function $F(\eta)$ given by \eqn{js2} after we replace $\tau$ by
$\eta$.
Indeed, with the help of \eqn{limi3}, we verify that,
for $\tau\to -\infty$ we approach the vacuum labeled
by $n=5$ in \eqn{vaa2} and \eqn{crrq}. In addition, it is easily seen that
for $\tau\to +\infty$ we approach the $n=0$ vacuum.
The shape of $\Om_1=\dots =\Om_5$ is similar to an anti-kink, whereas that of
$\Om_6=\Om_7$ is similar to a lump.

\section{7D self-dual YM with $G_2$ invariance}
\setcounter{equation}{0}

In this section we consider the self-dual YM equations in seven dimensions
in the case that their group of invariance is the $G_2$ subgroup
of the rotation group
$SO(7)$. As noted in \cite{Corrigan} this can be
obtained from the $Spin(7)$ invariant system \eqn{sellf} (with $\l=\ha$)
if we set to zero all the components of $F_{\a\b}$ that have
one subscript equal to 8. Indeed, then \eqn{sell1} gives a total of 7
conditions leaving the 21 equations in \eqn{sell2}, appropriate for the
7-dimensional system which reads\footnote{In this way we restrict to the
${\bf 14}$ of $G_2$. In complete analogy with the case of weak holonomy
metrics investigated recently in \cite{BDS}, this can be relaxed as we show
in subsection 6.1 below, leading to massive theories.}
\be
F_{ab}=\ha \psi_{abcd} F_{cd}\ , \qq a=1,2,\dots , 7\ .
\label{g2ll}
\ee
It is useful to split the index $a=(7,i,\hi)$ with $i=1,2,3$ and $\hi=i+3$
and represent the octonionic structure constants and the 4-index totally
antisymmetric tensor as
\ba
&&\psi_{ijk}=
\e_{ijk}\ ,\qq \psi_{i\hj\hk}= -\e_{ijk}\ , \qq \psi_{7i\hj}=\d_{ij}\ ,
\nonumber\\
&&\psi_{7ij\hk}=\e_{ijk}\ ,\qq \psi_{7\hi \hj \hk}=-\e_{ijk}\ , \qq
\psi_{ij\hat m \hat n}=\d_{im}\d_{jn}-\d_{in}\d_{jm}\ .
\ea
Similarly to the $Spin(7)$ invariant case, we can show that
a solution to a system of 7 equations suffices to construct a solution
of the entire set of self-dual equations \eqn{g2ll}. These are the 6
equations arising after choosing for the index $b=7$ in \eqn{g2ll}
\ba
&& F_{\hi 7}=\ha \psi_{\hi 7ab}F_{ab}= \ha \e_{ijk}(F_{\hj\hk}-F_{jk})\  ,
\nonumber\\
&& F_{i7}=\ha \psi_{i7ab}F_{ab}=- \e_{ijk} F_{j\hk}\
\label{iiin}
\ea
and in addition, the single equation
\be
F_{i\hi}=0\ ,
\label{iin1}
\ee
which is the only independent one from \eqn{iiin}
when the indices in the left hand side in
\eqn{g2ll} restrict to the values $a,b=1,2,\dots , 6$.
In order to reduce into one-dimensional systems we make the gauge
choice $A_7=0$ and we seek solutions where the remaining fields
$A_i$ and $\hat A_i \equiv A_{\hi+3}$ depend only on $x^7\equiv \tau$.
Then we obtain the system
\ba
{d\hat A_i\ov d\tau} & = & \ha \e_{ijk} ([\hat A_j,\hat A_k]-[ A_j, A_k])\ ,
\nonumber\\
{d A_i\ov d\tau} & = & -\e_{ijk} [ A_j,\hat A_k]\ ,
\label{aha}
\ea
with solutions subject to the constraint
\be
[A_i,A_\hi]=0\ ,
\label{cooon1}
\ee
which follows from \eqn{iin1}.
An equivalent useful complex form is
\be
{d\bar S_i\ov d\tau}=\ha  \e_{ijk}[S_j,S_k]\ ,
\qq
S_j = \hat A_j + i A_j\ ,
\label{ccom}
\ee
and $\bar S_i$ denotes its complex conjugation. The constraint becomes
\be
[S_i,\bar S_i]=0\ .
\label{cooon2}
\ee
Note that
this system is a complex extension of the Namh's system in \eqn{nahmm}.
It reduces to that in two cases: if $A_i=0, \ i=1,2,3$ or if
$A_i=\sqrt{3} \hat A_i, \ i=1,2,3$.

\bs
\no
\underline{\bf Reduction 1}: A quite general ansatz for metrics with
$G_2$ holonomy that preserve an $SU(2)\times SU(2)\times Z_2$ symmetry
was made in \cite{Brandhu} and \cite{cvBra}. The $G_2$ holonomy constraints
for the closure and co-closure of the associative three-form
result into a six-dimensional first order system of equations to which a
special solution having an extra $U(1)$ symmetry was found in \cite{Brandhu}.
Since there is no systematic study of this system,
we hope that making contact with solutions of self-dual YM will
give some new insight in this direction as well.
Hence, we seek solutions of \eqn{ccom} that
preserve at least an $SU(2)\times SU(2)\times Z_2$ symmetry.

Let us introduce two commuting sets of Pauli
matrices $\{\s_i\}$ and $\{\S_i\}$ with $i=1,2,3$ obeying
\be
[\s_i,\s_j]=2 i \e_{ijk} \s_k\ , \qq \tr(\s_i\s_j)=2 \d_{ij}\
\ee
and similarly for $\S_i$. Our
$SU(2)\times SU(2)\times Z_2$ invariant ansatz for the gauge fields is
\be
A_j(\tau)= -{i\ov 2} \om_j(\tau) (\s_j-\S_j) \ ,\qq
\hat A_j(\tau)=-{i\ov 2} \hat\om_j(\tau) (\s_j+\S_j) \ ,
\label{aannn}
\ee
where $\om_j$, $\hat \om_j$ are 6 functions to be determined.
The equivalently complex form is
\be
S_j(\tau)= -{i\ov 2} s_j(\tau) \s_j -{i\ov 2} \bar s_j(\tau) \S_j\ ,
\qq  s_j= \hat\om_j + i \om_j\ .
\ee
Our reduction is similar to the reduction of the Nahm system \eqn{nahmm}
to the Lagrange system \eqn{eell} which was considered in
\cite{Chakravarty,Takhtajan}.
We find that \eqn{ccom} reduce to
\be
{d\bar s_1\ov d\tau}
 = s_2 s_3\ ,
\qq ({\rm and\ cyclic\ perms.}) \ ,
\label{euul}
\ee
whereas the constraint \eqn{cooon2} is trivially satisfied.
In terms of the real components we have
\be
 {d\hat\om_1\ov d\tau} =\hat \om_2 \hat\om_3-\om_2\om_3\ ,\qq
{d\om_1\ov d\tau} =  -\hat\om_2 \om_3- \om_2 \hat\om_3\ ,
\qq ({\rm and\ cyclic\ perms.}) \ .
\ee
We emphasize that this system cannot be obtained as a particular case of
\eqn{om12} since the underline group structure is quite different.
Nevertheless, similarly to \eqn{om12},
our system is also a gradient flow with the non-vanishing
metric components, prepotential and potential given by
\ba
&& g_{s_i \bar s_j}= \d_{ij} \ ,
\qq W = s_1 s_2 s_3 + \bar s_1\bar s_2\bar s_3\ ,
\nonumber\\
&& V=|s_1s_2|^2+ |s_1s_3|^2+ |s_2s_3|^2\ .
\ea

The system \eqn{euul} is the complex extension of the 3-dimensional
Lagrange or Euler system \eqn{eell}
and it has an obvious symmetry under cyclic permutation in $1,2,3$.
However, there is a
less obvious discrete symmetry present only in this case which acts as
\be
\tau\to -\tau \ ,\qq
(\om_1,\om_2,\om_3)\to (\hat\om_1,\hat\om_2,-\om_3)\ ,\qq
(\hat\om_1,\hat\om_2,\hat\om_3)\to (\om_1,\om_2,\hat\om_3)\ .
\label{diiss}
\ee
Equivalent representations are obtained by cyclic permutation in the
indices 1, 2 and 3. This discrete symmetry originates from the automorphism
of the octonionic algebra \eqn{occtt} under the discrete transformation
obtained from \eqn{diiss}
if we replace the $\om_i$'s by the $e_i$'s, the $\hat \om_i$'s
by the $e_{i+3}$'s
and $\tau$ by $e_7$, as it can be readily verified.\footnote{A similar
discrete symmetry leaves invariant the full first order system for
6 functions of \cite{Brandhu,cvBra} obtained in the construction of
$G_2$ holonomy metrics having an $SU(2)\times SU(2)\times Z_2$ invariance
(joint work with I. Bakas).
The reason for this similarity is that,
the system of \cite{Brandhu,cvBra} can also be obtained
from imposing the self-duality condition on the
spin connection $\om_{ab}=\ha \psi_{abcd} \om_{cd}$ \cite{BDS}.}
In order to, at least partially, integrate the system \eqn{euul} it is
useful to recognize the constants of motion. It can be easily seen that
\ba
&&I_1=|s_1|^2-|s_2|^2=\om_1^2+\hat\om_1^2-\om_2^2-\hat\om_2^2\ ,
\nonumber\\
&&I_2=|s_2|^2-|s_3|^2=\om_2^2+\hat\om_2^2-\om_3^2-\hat\om_3^2\ ,
\label{coo12}
\ea
and
\be
I_3={i\ov 2} (s_1s_2s_3-\bar s_1 \bar s_2 \bar s_3)=
\om_1\om_2\om_3 -\om_1\hat\om_2\hat\om_3
-\om_2\hat\om_3\hat\om_1 - \om_3\hat\om_1\hat\om_2\ ,
\label{coo3}
\ee
are indeed constants of motion.
The first two are simply complex extensions of the
two independent integrals of motion of the 3-dimensional Euler top case.
We were unable to find additional independent constants of motion towards
further integrating the system \eqn{euul}.
In that respect it is important to
investigate if it admits a Lax pair representation as it is
the case for the usual Lagrange system \eqn{eell} (see, for instance,
\cite{Takhtajan}). Nevertheless, we remark that in the case with $\om_1=\om_2$
and $\hat \om_1=\hat \om_2$, where our ansatz \eqn{aannn} develops an extra
$U(1)$ invariance, we have 2 independent constants of motion $I_2$ and
$I_3$ which can be used to fully integrate the system for the
remaining 4 functions.\footnote{There is an analogy for this in the
6 function system of \cite{Brandhu,cvBra} describing the
$G_2$ holonomy metrics preserving an $SU(2)\times SU(2)\times Z_2$ symmetry.
In the case that an extra $U(1)$ symmetry factor
develops the system reduces to one for 4
functions and a special solution to it was found \cite{Brandhu}.}
Indeed, if we change variables
as $s_j = e^{i \phi_j}r_j$ we can rewrite the system \eqn{euul} as
\ba
{dr_1\ov d\tau} & =& r_2 r_3 \cos(\phi_1+\phi_2+\phi_3)\ ,\qq
({\rm and\ cyclic\ perms.})\ ,
\nonumber\\
{d\phi_1\ov d\tau}& =& -{r_2 r_3\ov r_1} \sin(\phi_1+\phi_2+\phi_3)\ ,\qq
({\rm and\ cyclic\ perms.})\ .
\ea
In the case that $\phi_1=\phi_2$ and $r_1=r_2$ we can easily show that
all functions are determined in terms of a single function $x(\tau)$
as
\ba
&& r_1^2=r_2^2=x+{I_2\ov 3}\ ,\qq r_3^2 = x-{2I_2\ov 3}\ ,
\nonumber\\
&& {d\phi_1\ov d\tau}={d\phi_2\ov d\tau}={I_3\ov x+I_2/3}\ ,\qq
{d\phi_3\ov d\tau}={I_3\ov x-2 I_2/3}\ .
\ea
The function $x(\tau)$ obeys the differential equation
\be
\left(d x\ov d \tau\right)^2
= 4 x^3 - {4\ov 9} I_2^2 x - 4I_3^2 - { 8 \ov 27} I_2^3\ ,
\ee
which is of the Weierstrass form and therefore its solution
is given in terms of the corresponding elliptic functions.

\bs
\no
\underline{\bf Reduction 2}: There is alternative reduction one can perform
that resembles that of
\cite{Chakravarty,Takhtajan} and their reduction of the Nahm system
\eqn{nahmm} to the Halphen system \eqn{eell1}.
In order to proceed we need to introduce
the group element  $g\in SU(2)$ and construct the components $L^j_\m$ of
the left invariant Maurer--Cartan 1-forms $L^j$ and the matrix $C_{ij}$ as
\be
L^j_\m=-i \tr(g\inv \del_\m g \s_j)\ ,\qq
C_{ij}=\ha \tr(\s_i g \s_j g\inv)\ ,
\ee
where $x^\m$, $\m=1,2,3$ represent the variables that we use to parametrize
the group element $g\in SU(2)$.
Some useful properties
\ba
&& \del_\m L^i_\n - \del_\n L^i_\m = 2 \e_{ijk} L^j_\m L^k_\n\ ,
\nonumber\\
&& C_{ik} C_{jk}= \d_{ij}\ , \qq C_{im}C_{jn}\e_{mnl}=\e_{ijk}C_{kl}\ .
\ea
Also we recall the left-invariant $SU(2)$ vector fields
$X_i$, $i=1,2,3$ satisfying
\be
[X_i,X_j]= -2 \e_{ijk}X_k \ ,\qq [X_k,C_{ij}]=2 C_{il} \e_{ljk} \ .
\ee
In terms of the inverses $L^\m_i$ of the $L^i_\m$'s they are can be
represented as the differential operators $X_i=L^\m_i \del_\m$.
Then the alternative
parametrization of the gauge fields is
\be
A_i(\tau)  =\ha \om_j(\tau) C_{ij} X_j\ , \qq \hat A_i(\tau)
=\ha \hat \om_j(\tau) C_{ij} X_j\ ,  \qq
S_i(\tau)=\ha s_j(\tau) C_{ij} X_j\ .
\label{annn2}
\ee
Then \eqn{ccom} gives rise to the system
\be
 {d\bar s_1\ov d\tau}
 = -s_2 s_3 + s_1(s_2+s_3)\ ,\qq ({\rm and\ cyclic\ perms.}) \ ,
\ee
which is a complex generalization of the Halphen system \eqn{eell1}.
One can also show that the constraint \eqn{cooon2} is respected by the
ansatz \eqn{annn2}.
For the real components we have
\ba
&&{d\hat\om_1\ov d\tau}= \om_2\om_3 -  \om_1(\om_2+
 \om_3)- \hat\om_2\hat\om_3 + \hat\om_1(\hat\om_2+\hat\om_3)\ ,
\quad ({\rm and\ cyclic\ perms.}) \ ,
\nonumber\\
&&{d\om_1\ov d\tau} = \hat\om_2\om_3 +\hat\om_3\om_2 - \hat\om_1(\om_2+
\om_3)-  \om_1(\hat\om_2+\hat\om_3)\ ,\quad ({\rm and\ cyclic\ perms.}) \ .
\label{ommh}
\ea
As previously this system is a gradient flow with non-vanishing metric
components, prepotential and potential given by
\ba
&& g_{s_i \bar s_j}= 1-2 \d_{ij} \ ,\qq
 W = s_1 s_2 s_3 + \bar s_1\bar s_2\bar s_3\ ,
\nonumber\\
&&V=|s_1 s_2 + s_1 s_3 + s_2 s_3|^2 -2 (|s_1s_2|^2+ |s_1s_3|^2+ |s_2s_3|^2)\ .
\ea
Note that the expression \eqn{coo3} is a constant of motion in this case
as well. However, we were not able to find additional constants of motion.
Also notice that, the discrete transformation \eqn{diiss} is no
longer a symmetry of the system \eqn{ommh}.

\subsection{The massive case}

We may choose in passing from the 8-dimensional
self-duality conditions \eqn{sell1} and \eqn{sell2} to the 7-dimensional
ones to keep a non-zero piece proportional to $\psi_{abc} F_{bc}$ transforming
as a ${\bf 7}$ of $G_2$. We only require that it is
proportional to the gauge field $A_a$, i.e. $F_{8a}=m A_a$.
This is in complete analogy with the weak holonomy metrics
(a notion introduced in \cite{gray}) that were recently investigated
in \cite{BDS}.
Then \eqn{g2ll} is modified as
\be
F_{ab}=\ha \psi_{abcd} F_{cd} -{m} \psi_{abc}A_c\ .
\ee
We emphasize that even in this case the seven conditions $F_{7i}=\dots$,
$F_{7\hi}=\dots$ and $F_{i\hi}=\dots$ suffice to satisfy the rest.
Then, by performing the usual reduction to one dimension we find that
the system \eqn{aha} and the constraint \eqn{cooon1} are modified as
\ba
{d\hat A_i\ov d\tau} & = & \ha \e_{ijk} ([\hat A_j,\hat A_k]-[ A_j, A_k])
+ m A_i \ ,
\nonumber\\
{d A_i\ov d\tau} & = & -\e_{ijk} [ A_j,\hat A_k] - m \hat A_i\ ,
\label{ahaw}
\ea
and
\be
[A_i,A_\hi]= -m A_7\ .
\label{coon1w}
\ee
This shows clearly that the effect of keeping the ${\bf 7}$ is to produce
a mass term.
However, we will not preceed in investigating further this system
in the present paper.

\section{Concluding remarks}

We have shown that certain reductions to one dimension of the $Spin(7)$
invariant 8-dimensional self-dual YM equations result into systems that
can be completely integrated using techniques from algebraic geometry. The
different inequivalent solutions are characterized by the genus
of certain Riemann surfaces.
We find it rather remarkable that the same system and its solution
arose in the constructions of domain wall solutions in gravitational
theories with scalars, corresponding to sectors of gauged supergravities
in diverse dimensions. These solutions were constructed before \cite{bs1,bs2}
in studies of the Coulomb branch of
SYM theories in the context of the AdS/CFT correspondence.
We have also considered reductions to one dimension
of the $G_2$ invariant 7-dimensional self-dual YM equations that
preserve an $SU(2)\times SU(2)\times Z_2$ symmetry. This system has
the same symmetries as a similar system that arose recently in the
construction of $G_2$ holonomy metrics in Euclidean 7-dimensional gravity
\cite{Brandhu,cvBra}.
It is very interesting to explore the possibility
that there is a change
of variables that maps one system to the other.
We have been able
to find three integrals of motion for the system \eqn{euul} and in fact
solve it in general for the case that there is an extra $U(1)$ symmetry.
Hence, if such a mapping exists, it will be advantageous towards
systematizing the search for new $G_2$
holonomy manifolds. We hope to report work along this direction in the
feature.

\bs\bs
\centerline{\bf Acknowledgements}

This research was supported by the European Union under
TMR-ERBFMRX-CT96-0045 and -0090, by the Swiss Office for Education and
Science, by the Swiss National Foundation and by the contract
HPRN-CT-2000-00122.


\newpage

\appendix

\setcounter{equation}{0}
\renewcommand{\theequation}{\thesection.\arabic{equation}}

\section{Octonionic identities}

The octonionic non-associative algebra is given by
\be
e_a\cdot e_b =- \d_{ab} e_0 + \psi_{abc} e_c\  ,\qq a,b,c=1,2,\dots , 7\ ,
\label{occtt}
\ee
where $e_0$ is the unit element and $\psi_{abc}$ are the octonionic
structure constants. In the standard basis \cite{Octref}
\be
\psi_{123}=\psi_{516}=\psi_{624}=\psi_{435}=\psi_{471}=\psi_{673}=
\psi_{572}=1\ .
\label{g2str1}
\ee
The 4-index totally antisymmetric tensor is defined as
\be
\psi_{abcd}={1\ov 3!}\e_{abcdefg} \psi_{efg}\
\ee
and in the standard basis is given by
\be
\psi_{1245}=\psi_{2671}=\psi_{3526}=\psi_{4273}=
\psi_{5764}=\psi_{6431}=\psi_{7531}=1\ .
\label{g2str2}
\ee

The tensors $\psi_{abc}$ and $\psi_{abcd}$
can be assembled in a single object $\Psi_{\a\b\g\d}$ of $SO(8)$ as:
\be
\a=(i,8)\ , \qq \Psi_{abc 8}= \psi_{abc}\ ,\qq \Psi_{abcd}=\psi_{abcd}\ .
\ee
Then
\be
\Psi_{\a\b\g\d}={\e\ov 4!}
\e_{\a\b\g\d\zeta\eta\s\kappa}\Psi^{\zeta\eta\s\kappa}\ ,\qq e=\pm 1\ .
\ee
For $\e=1$ it is selfdual and for $\e=-1$ antiselfdual.

The basic identity is\footnote{If $\Psi_{\a\b\g\d}$ was replaced by
$\e_{\a\b\g\d}$ in four-dimensions, only the first line below should be kept.}
\ba
 \Psi_{\a\b\g\d}\Psi^{\zeta\eta\s\d} & =  &
\d^\s_\g (\d^{\zeta}_\a \d^\eta_\beta - \d^{\zeta}_\b \d^\eta_\a)  +
\d^\eta_\g (\d^{\s}_\a \d^\zeta_\beta - \d_{\b}^\s \d^\zeta_\a)  +
\d^\zeta_\g (\d^{\eta}_\a \d^\s_\beta - \d^{\eta}_\b \d^\s_\a)
\nonumber\\
&&
- \e (\Psi_{\a\b}{}^{\zeta\eta} \d^\s_\g + \Psi_{\a\b}{}^{\s\zeta} \d^\eta_\g
+ \Psi_{\a\b}{}^{\eta\s} \d^\zeta_\g )
\nonumber\\
&&
- \e (\Psi_{\g\a}{}^{\zeta\eta} \d^\s_\b + \Psi_{\g\a}{}^{\s\zeta} \d^\eta_\b
+ \Psi_{\g\a}{}^{\eta\s} \d^\zeta_\b )
\nonumber\\
&&
- \e (\Psi_{\b\g}{}^{\zeta\eta} \d^\s_\a + \Psi_{\b\g}{}^{\s\zeta} \d^\eta_\a
+ \Psi_{\b\g}{}^{\eta\s} \d^\zeta_\a )\ .
\ea
From this we derive several other identities

\be
\Psi_{\a\b\zeta\eta}\Psi^{\g\d\zeta\eta}  =
6(\d^{\g}_\a \d^\d_\beta - \d^{\g}_\b \d^\d_\a)  -4 \e \Psi_{\a\b}{}^{\g\d}\ .
\ee
and
\ba
\psi_{abf} \psi^{cdef}& =&  -\e \psi_{ade} \d^c_b - \e \psi_{acd} \d^e_b
-\e \psi_{aec} \d^d_b + \e \psi_{bde} \d^c_a + \e \psi_{bcd} \d^e_a
+ \e \psi_{bec} \d^d_a\ ,
\nonumber\\
\psi_{abe}\psi^{cde}& = & \d^{c}_a \d^d_b - \d^{c}_b\d^d_a  - \e
\psi_{ab}{}^{cd}\ ,
\\
\psi_{abef}\psi^{cdef}& = & 4(\d^{c}_a \d^d_b - \d^{c}_b\d^d_a)  -2 \e
\psi_{ab}{}^{cd}\ ,
\nonumber\\
\psi_{abde}\psi^{cde} & =&  -4 \e \psi_{ab}{}^c\ .
\ea

We can define an adjoint-like representation for seven-dimensional matrices
as
\be
\psi_a : \qq   (\psi_a)_{bc}=\psi_{abc}\ .
\label{addj}
\ee
Then, using the above properties we find that
\be
[\psi_a,\psi_b]_{cd} = \d_{ad}\d_{bc} -\d_{ac}\d_{bd} - 2\e \psi_{abcd}\ .
\label{addj1}\ee
and
\be
\psi_{abc} [\psi_b,\psi_c] =6 \psi_a \ ,\qq {\rm Tr}(\psi_a\psi_b)=
-6 \d_{ab} \ , \qq  {\rm Tr}([\psi_a,\psi_b]\psi_c)= -6 \psi_{abc}\ .
\label{addj2}
\ee



\end{document}